\documentclass[sigconf,nonacm]{acmart}

\AtBeginDocument{%
  \providecommand\BibTeX{{%
    \normalfont B\kern-0.5em{\scshape i\kern-0.25em b}\kern-0.8em\TeX}}}

\usepackage{amsfonts,amsmath,amssymb,amsthm,bm}

\begin{document}

\title[]{Generative Adversarial Network for Wireless Signal Spoofing
}

\author{Yi Shi}
\affiliation{%
  \institution{Intelligent Automation, Inc.}
  \city{Rockville}
  \state{MD}
  \postcode{20855}}
\email{yshi@i-a-i.com}

\author{Kemal Davaslioglu}
\affiliation{%
  \institution{Intelligent Automation, Inc.}
  \city{Rockville}
  \state{MD}
  \postcode{20855}}
\email{kdavaslioglu@i-a-i.com}

\author{Yalin E. Sagduyu}
\affiliation{%
  \institution{Intelligent Automation, Inc.}
  \city{Rockville}
  \state{MD}
  \postcode{20855}}
\email{ysagduyu@i-a-i.com}


\renewcommand{\shortauthors}{ }

\vspace{1in}
\begin{abstract}
The paper presents a novel approach of spoofing wireless signals by
using a general adversarial network (GAN) to generate and transmit synthetic
signals that cannot be reliably distinguished from intended signals. It is of
paramount importance to authenticate wireless signals at the PHY layer
before they proceed through the receiver chain. For that purpose, various waveform, channel, and radio hardware features that are inherent to original wireless signals need to be captured. In the meantime, adversaries become sophisticated with the cognitive radio capability to record, analyze, and manipulate signals before spoofing. Building upon deep learning techniques, this paper introduces a spoofing attack by an adversary pair of a transmitter and a receiver that assume
the generator and discriminator roles in the GAN and play a minimax game to
generate the best spoofing signals that aim to fool the best trained
defense mechanism. The output of this approach is two-fold. From the
attacker point of view, a deep learning-based spoofing mechanism is
trained to potentially fool a defense mechanism such as RF
fingerprinting. From the defender point of view, a deep learning-based
defense mechanism is trained against potential spoofing attacks when an adversary pair
of a transmitter and a receiver cooperates. The probability that the spoofing signal is misclassified as the intended signal is measured for random signal, replay, and GAN-based spoofing attacks. Results show that the GAN-based spoofing attack provides a major increase in the success probability of wireless signal spoofing even when a deep learning classifier is used as the defense.
\end{abstract}

%

\keywords{Adversarial machine learning, deep learning, general adversarial network (GAN), spoofing attack.}


\maketitle

\section{Introduction}
Wireless communications is susceptible to adversaries due to the open and shared nature of wireless medium. Among various wireless attacks, the spoofing attack is launched by an adversary that aims to mimic a legitimate user in its transmissions. This attack may serve different purposes including primary user emulation (PUE) in cognitive radio networks, passing through signal authentication systems, and intrusion into protected networks. One common approach for wireless signal spoofing is recording a legitimate user's transmission and replaying the signal later by potentially adjusting the transmit power. While this approach can represent various features in the signal at a high level, it may fall short of reliably mimicking combined waveform, channel, and device effects. In this context, machine learning provides automated means to authenticate signals by analyzing wireless signals and identifying anomalies. Enabled by recent advances in computational resources, deep learning can effectively process raw spectrum data and operate on latent representations, while analyzing high-dimensional spectrum dynamics that feature-based machine learning algorithms fail to achieve.

Deep learning finds rich applications in wireless communications, including spectrum sensing \cite{Kemal2018} and  modulation recognition \cite{OShea2016}. On the other hand, the adversary can also apply deep learning to launch wireless attacks. For example, an adversary may try to learn the underlying transmit behavior by training a deep neural network and effectively jam data transmissions \cite{Yi18:Jamming}. There are various security concerns regarding the safe use of machine learning algorithms.
\emph{Adversarial machine learning} \cite{AMLbook, Sec2} studies learning in the presence of adversaries and aims to enable safe adoption of machine learning to the emerging applications such as wireless communications.
For example, the adversary may manipulate input data into a machine learning classifier by jamming the sensing period \cite{Yi18:Poisoning}. Similarly, an adversary may launch an evasion attack by manipulating signals over the air so as to force a receiver in making wrong signal classification decisions \cite{LarssonAML, Headley19, Deniz19, Silvija19}.

In this paper, we introduce a \emph{spoofing attack} motivated by adversarial machine learning. In particular, consider training a \emph{generative adversarial network} (GAN) to spoof wireless signals as if they originate from intended users (legitimate or higher priority users such as primary users). GAN has been extensively applied to other domains such as computer vision and text analytics to generate synthetic data that is statistically similar to real data \cite{Goodfellow2014}. Recently, there have been efforts to apply GAN to wireless communications. The focus has been to augment the training data sets such as those used to train classifiers for spectrum sensing \cite{Kemal2018} and jamming \cite{Tugba2018}. In this paper, our goal is to train a GAN from an adversarial point of view to spoof wireless signals that cannot be reliably discriminated from intended signals. Different from applications in other domains such as computer vision, data in wireless medium is received through channel and (transmitter and receiver) hardware effects, and depends on transmitter-receiver positions that all need to be matched by the GAN. In this context, a receiver (assuming the role of a defender) aims to classify signal sources as intended user or not. This classification can be done by using a deep learning based classifier to analyze spectrum sensing results. Then the adversary launches a spoofing attack such that the classifier at the receiver incorrectly authenticates its transmissions as intended.

As a starting point, we show that a deep learning-based classifier can distinguish signals of an intended transmitter $T$ from other randomly generated signals.
Given that each device introduces its own phase shift and each channel has its own propagation gain and phase shift, such a classifier can be built at a receiver $R$ by collecting spectrum sensing results for signals from $T$ and other signals, and processing them to obtain a number of features  for each sample.
These features and labels (from $T$ or not) form the training data to build a classifier by deep learning.
We show that such a classifier can successfully distinguish different signals.
In particular, we can regard signals from other transmitter as a naive spoofing attack with random signals.
This naive attack, as we expected, does not perform well.
Its success probability (the probability that signals from other transmitter is classified as from $T$) is limited to only $7.89\%$.

We then study the replay attack, where an adversary transmitter $A_T$ amplifies and forwards the previously received signal from $T$.
This attack keeps some pattern of $T$ (but not the entire), and thus is better than using random signals.
We show that the success probability increases to $36.2\%$ against a defender that uses a deep learning classifier.
However, this probability is still much less than $50\%$, i.e., most of replay-based spoofing attacks based on amplifying and forwarding signals are still not successful.

To launch a successful spoofing attack, the adversary transmitter $A_T$ needs to generate signals such that  signals received from $A_T$ are statistically similar to signals received from $T$.
The challenge is that $A_T$ does not have any knowledge on $T$'s waveform (in our case, characterized by the modulation scheme), its phase shift, as well as the channel between $T$ and $R$.
In this paper, we introduce a GAN-based approach to capture all these effects from observed signals and generate spoofing signals without the need of prior knowledge (that may not be available at all due to unpredictable spectrum dynamics). The transmitter and its surrogate receiver (used only for training) are trained offline as the generator-discriminator pair of the GAN to generate the best spoofing signals that aim to fool the best-trained defense mechanism. The output is a signal generator that creates wireless signals with fake signatures to spoof signals. This generator is trained as a deep neural network to fool the optimized discriminator that is trained as another deep neural network. 

During the training, $A_T$ provides some flag to label its transmissions, and thus $A_R$ knows the true label.
$A_R$ processes received signals, obtains features, and builds a discriminator to classify signals as from $T$ or not.
The classification results are sent back to $A_T$ as feedback.
Then $A_T$ updates its generator to generate better synthetic data, namely to increase classification error probability at $A_R$.
Thus, $A_T$ and $A_R$ play a minimax game, which forms a GAN to improve both generator and discriminator.
Once GAN converges, the generator at $A_T$ can generate high fidelity synthetic data (similar to real data) for spoofing attack. This approach inherently captures all waveform, channel, and device effects jointly.
As a result, the success probability of spoofing attack increases to $76.2\%$, when the GAN-based approach is used.

The rest of the paper is organized as follows.
Section~\ref{sec:related} discusses related work.
Section~\ref{sec:scenario} describes the system model.
Section~\ref{sec:classifier} describes the pre-trained classifier to detect intended transmissions.
Section~\ref{sec:adversary} describes and compares the replay and GAN-based spoofing attacks.
Section~\ref{sec:conclusion} concludes the paper.

\section{Related Work}
\label{sec:related}

There are different types of attacks on wireless communications in the literature \cite{Clancy08:CogSec}. In particular, attacks on spectrum sensing include spectrum sensing data falsification (SSDF) attack \cite{Penna, Sagduyu2014}, 
primary user emulation (PUE) attack \cite{PUE}, eavesdropping \cite{Zou15:eavesdropping}, and noncooperation \cite{Sagduyu09:noncoop}.
Attacks on data transmission include jamming \cite{Sagduyu118:satellite} in form of a denial-of-service (DoS) attack \cite{DoS} with different levels of prior information \cite{Sagduyu11:jamming}.
There are also attacks on higher layers, e.g., attacks on routing in the network layer \cite{Lu2017} and network flow inference attacks \cite{LuCliff2017}.
Defense methods were developed to address these attacks.
For example, an adaptive, jamming-resistant spectrum access protocol was proposed in \cite{JamRes} for cognitive radio ad hoc networks, where there are multiple channels that the secondary users can utilize. Jamming games between a cognitive user and a smart jammer was considered in \cite{UserCentric}, where they individually determine their transmit powers.

Launching and detecting spoofing attacks have been extensively studied \cite{Lichtman16:spoofing, Gai17:spoofing, Chen07:spoofing, Sheng08:spoofing, Sajjad18:spoofing}.
Spoofing along with other attacks such as jamming and sniffing were assessed in \cite{Lichtman16:spoofing}
A spoofing attack was  designed in \cite{Gai17:spoofing} using optimal power distribution.
In this paper, we optimize both power and phase shift for spoofing attack.
Received signal strength (RSS) was used to in \cite{Chen07:spoofing,Sheng08:spoofing} detect spoofing attack. In this paper, we use raw spectrum sensing (I/Q) samples.
Recently, deep learning was also applied to detect spoofing attacks, e.g., CNN was used in \cite{Sajjad18:spoofing}, while the use of deep learning in this paper is to optimize launching spoofing attacks.

Wireless security finds rich applications of deep learning. Deep learning was applied to authenticate signals \cite{Saad2018}, detect and classify jammers of different types \cite{Poor2018, Wu2017}, 
and control communications to mitigate jamming effects \cite{Poor2018, UserCentric}.
Jammers typically do not use machine learning techniques, e.g., \cite{Poor2018, 
EnergyHarvestingCN}. Recently, there have been efforts to build deep learning-based jammers \cite{Yi18:Poisoning, Yi18:Jamming}.
Using wireless sensors, deep learning was also used to infer private information in analogy to exploratory attacks \cite{Liang18}. From a different perspective, GAN was applied to model wireless communication channels, e.g., \cite{Yang19:channel, Oshea18:channel, Oshea18:channel2}.
In this paper, we use GAN to generate synthetic spoofing signals, which need to model not only channel effects but also device related effects such as phase shift as well as relative positions of transmitters and receivers from both attacker and defender sides.

Adversarial deep learning was applied to launch evasion attacks by adding perturbations to the received signals and  manipulating the input to the machine learning algorithm. \cite{LarssonAML, Headley19, Deniz19, Silvija19} considered evasion attacks against modulation classifiers.
In this paper, we consider a spoofing attack with the same final goal as the evasion attack. However, the classifier to fool in this paper does not only use modulation but also channel effects, device related effects such as phase shifts, and relative positions of transmitters and receivers of both attacker and defender. As a baseline, we consider the replay (amplify and forward) attack \cite{Kinnunen17:replay, Hoehn16:replay} as a method of spoofing.
However, although replay attack can keep some features in original signals, it is not very effective.
The limits of replaying signals as the spoofing attack were assessed in \cite{Kinnunen17:replay}.
Also, schemes to detect replay attacks were reported in \cite{Hoehn16:replay}.
In this paper, our results show that GAN-based spoofing significantly outperforms the replay attack.

\section{System Model}
\label{sec:scenario}

We consider a wireless communication environment with intended users (such as legitimate or high-priority) and others (such as the adversary).
We assume that a pre-trained deep learning-based classifier is used at a receiver to predict whether a transmission is from an intended one, or not.

\begin{figure}
	\centering
	\includegraphics[width=\columnwidth]{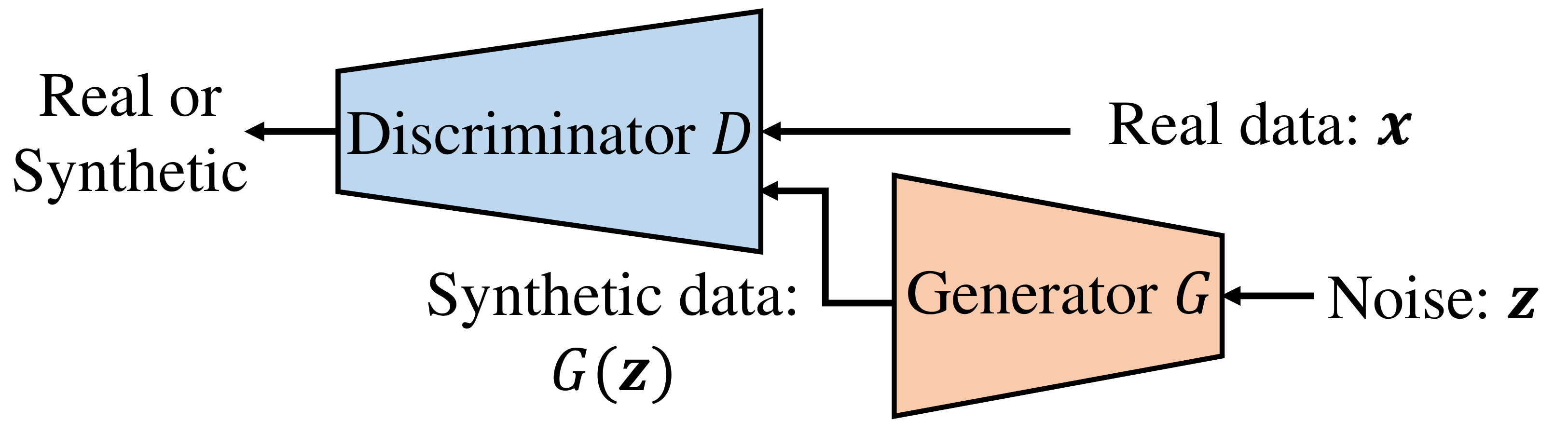}
	\caption{GAN structure.}
	\label{fig:gan}
\end{figure}

The adversary aims to launch a spoofing attack such that its transmissions are classified as an intended one.
Due to unique device properties (such as phase shift) and communication channel properties (such as channel gain), we show that random signal transmission by an adversary can be easily detected as an unintended transmission.
Thus, the adversary needs to learn the pattern of intended transmissions and generate its transmissions following the same pattern for spoofing attack.
Two adversaries collaborate for this purpose and act as a transmitter-receiver pair to train a GAN (see Fig.~\ref{fig:gan}). In particular, a generator is trained at the transmitter and a discriminator is trained at the receiver. In the spoofing attack (namely, test phase), only the generator at the transmitter is used.

\begin{figure}
	\centering
	\includegraphics[width=\columnwidth]{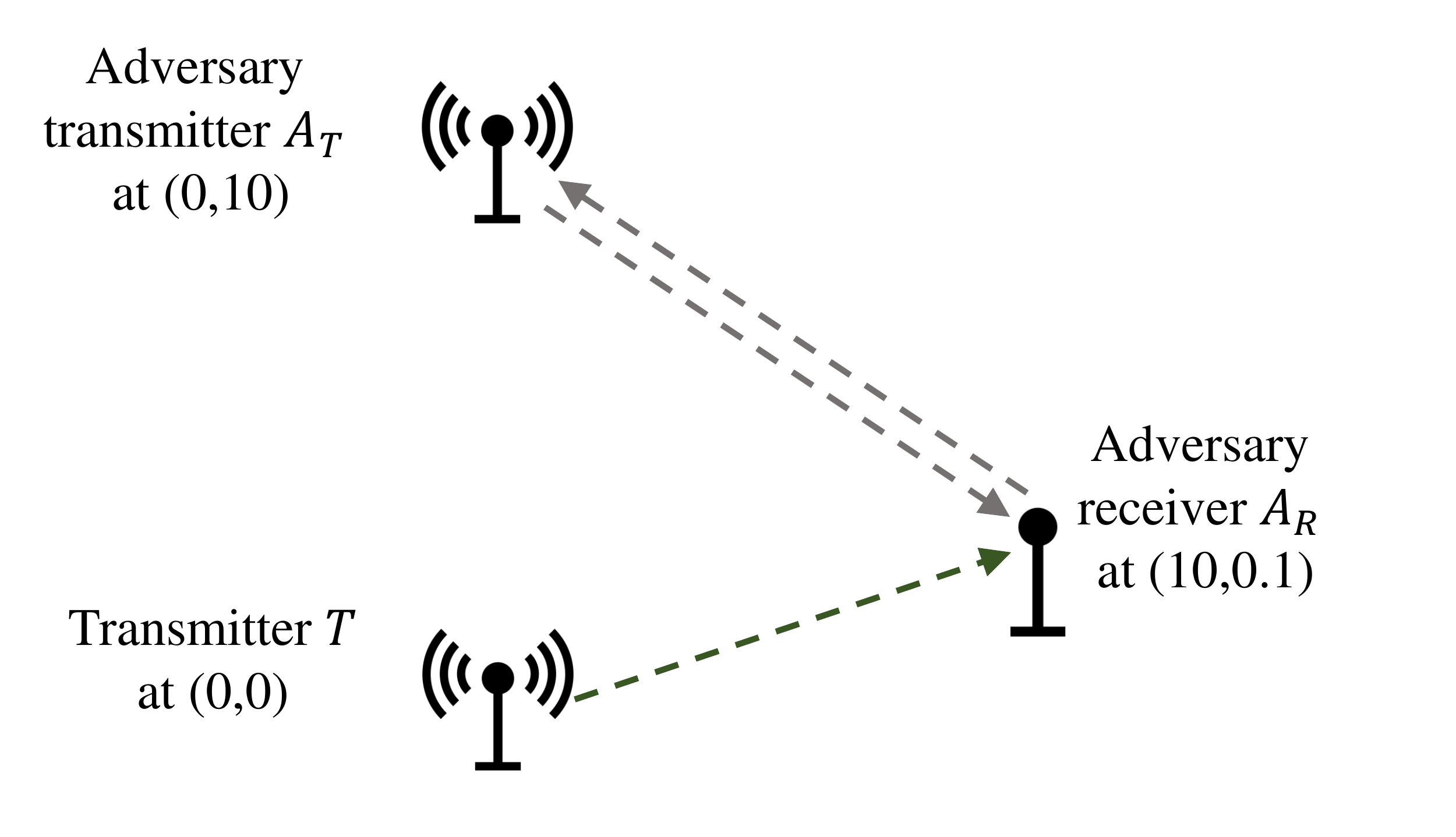}
	\caption{Network topology during the training process for spoofing attack.}
	\label{fig:topology1}
\end{figure}

\begin{figure}
	\centering
	\includegraphics[width=\columnwidth]{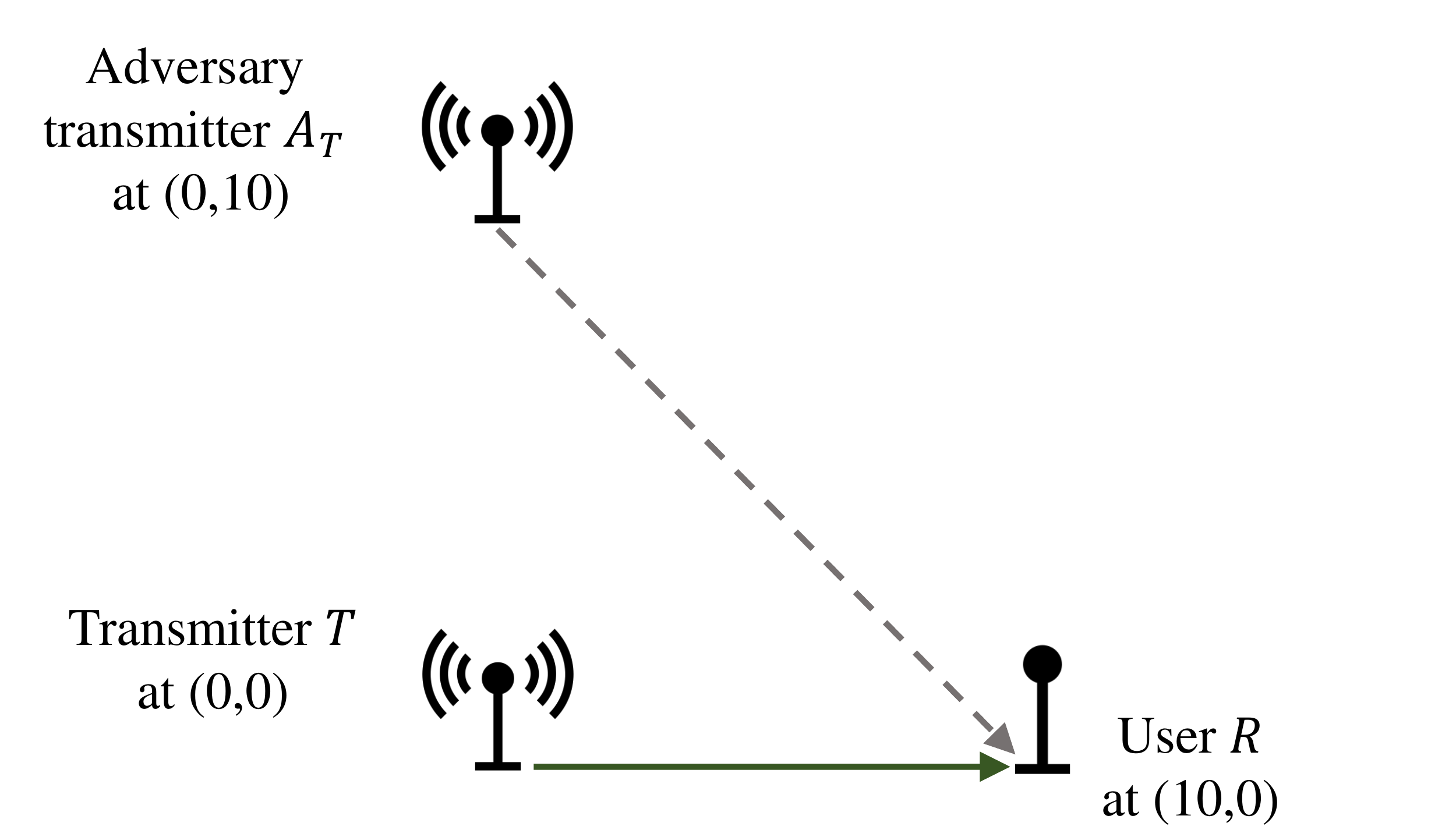}
	\caption{Network topology during the spoofing attack.}
	\label{fig:topology2}
\end{figure}

In this paper, we consider the scenario that there is an intended transmitter $T$, a receiver $R$ (that classifies if received signals are from $T$, or not), an adversary transmitter $A_T$, and an adversary receiver $A_R$.
$T$ transmits with power $P=1000$.
There is a device-related phase shift for $T$, which is unknown to adversaries $A_T$ and $A_R$.
We assume a Rayleigh channel between any two nodes.
$R$ has a deep learning-based classifier to classify whether signal is from $T$ or other transmitters, e.g., $A_T$.
We consider three methods for spoofing attacks.
\begin{itemize}
\item [1.] \emph{Random signal attack:} $A_T$ performs random transmissions with power $P$.

\item [2.] \emph{Replay attack:} $A_T$ amplifies and forwards previously received signal from $T$. Since $A_T$ does not have any knowledge on channel gains, it cannot optimally tune its power. Thus, $A_T$ uses power $P$ to amplify signals.

\item [3.] \emph{GAN-based spoofing attack:} $A_T$ generates synthetic signals using GAN. The transmission power is up to $P$.
\end{itemize}
The generation process of the GAN-based spoofing attack is as follows.
The adversary receiver $A_R$ is placed close to $R$ such that channel from $T$ (or $A_T$) to $A_R$ is similar to channel $T$ (or $A_T$) to $R$.
Thus, $A_R$ can receive similar signals as $R$ and if $A_R$ cannot distinguish signals from $T$ and $A_T$, $R$ cannot either.
$A_T$ makes its transmissions with a flag such that $A_R$ knows these signals are from $A_T$ (i.e., the true label).
$A_R$ builds a discriminator to classify signals from $T$ or $A_T$, and transmits the classification results to $A_T$ as feedback. Then $A_T$ builds a generator to improve its transmitted signals such that these signals are more similar to $T$'s signals in terms of resulting in larger classification error at $A_R$.
This process continues several rounds until convergence.
In this setting, $A_T$ and $A_R$ play a minimax game, which is exactly the GAN process (see Fig.~\ref{fig:topology1}).
When the GAN converges, the generator at $A_T$ should be able to generate synthetic signals very similar to $T$'s signals, which are then used for spoofing attack (see Fig.~\ref{fig:topology2}).

The advantage of this attack is that the adversary does not need to assume any prior knowledge on $T$, which will be learned by $A_R$.
Moreover, the adversary does not need to learn channel effect explicitly.
Instead, channel effects such as phase shift and propagation gain are learned implicitly through collaboration of $A_T$ with $A_R$.

\section{The Classifier for Signal Authentication}
\label{sec:classifier}

In this section, we describe the pre-trained classifier at $R$ that needs to identify whether a transmission is from $T$ or not. We assume that $T$ transmits data using the QPSK modulation.
Note that other modulations can also be used without changing the algorithms (classifiers) developed in this paper.
The transmit power is $P$ and the wireless channels are Rayleigh channel.
$R$ uses limited sensing data, say $8$ bits of data.
Under QPSK, there are four possible modulated signals, where each signal may have a different phase shift for $2$ bits.
In addition, each device has its own phase shift.
Denote $\theta_T$ as the phase shift of $T$, which is added on the QPSK signal's phase shift.
Note that other settings on the number of bits in sensing data and modulation will not affect the developed algorithms.

$R$ trains a deep neural network classifier to analyze spectrum sensing results and identifies whether a transmission is from $T$ or not.
This classifier is pre-trained using many samples with labels on whether a transmission is from $T$ or not.
Each sample has four signals and each received signal is sampled $100$ times.
Thus, there are $400$ features for each sample.
As an example, for two bits $0$ and $0$, QPSK determines a phase shift $\frac{\pi}{4}$ for coded signal.
Adding $\theta_T$ and a random channel phase shift $\theta_{T R}$ under the Rayleigh model, the received signal has phase shift $\frac{\pi}{4} + \theta_T + \theta_{T R}$.
The $k$-th sample point, $0 \le k < 100$, has phase shift $\frac{\pi}{4} + \theta_T + \theta_{T R} + \frac{k \pi}{50}$.
The received power is $g_{T R} P$, where $g_{T R}$ is a random channel gain under the Rayleigh model.
The mean value of channel gain is $d^{-2}$, where $d$ is the distance between a transmitter and a receiver.
Then the $k$-th sampled data is
\begin{eqnarray}
d_{TR}^k = g_{TR} P e^{j (\frac{\pi}{4} + \theta_T + \theta_{TR} + \frac{k \pi}{50})} \; .
\end{eqnarray}
During the training, $T$ may send a flag to indicate its transmissions and this flag is used to label samples.
After observing a certain period of time, $R$ collects a number of samples with labels to be used as training data to build a deep learning classifier.

Once a classifier is built, $R$ uses it to predict signal labels (`$T$' or `not $T$').
For this algorithm, there may be two types of errors:
\begin{itemize}
\item \emph{Misdetection}. The signal is from $T$ but it is detected as from other transmitters.

\item \emph{False alarm}. The signal is from other transmitters but it is detected as from $T$.
\end{itemize}
$R$ aims to minimize the probability of both errors.
Denote $e_{MD}$ and $e_{FA}$ as the probabilities of misdetection and false alarm at $R$, respectively.
Then the objective is to minimize $\max\{ e_{MD}, e_{FA} \}$.
For a given test data with $n$ samples and $n_T$ as the number of samples with signals from $T$, denote $n_{MD}$ as the number of misdetections and $n_{FA}$ as the number of false alarms.
These error probabilities are calculated by
\begin{eqnarray} e_{MD} = \frac{n_{MD}}{n_T} ,\; \:\:  e_{FA} = \frac{n_{FA}}{n-n_T} \; .
\end{eqnarray}

\begin{figure}
	\centering
	\includegraphics[width=\columnwidth]{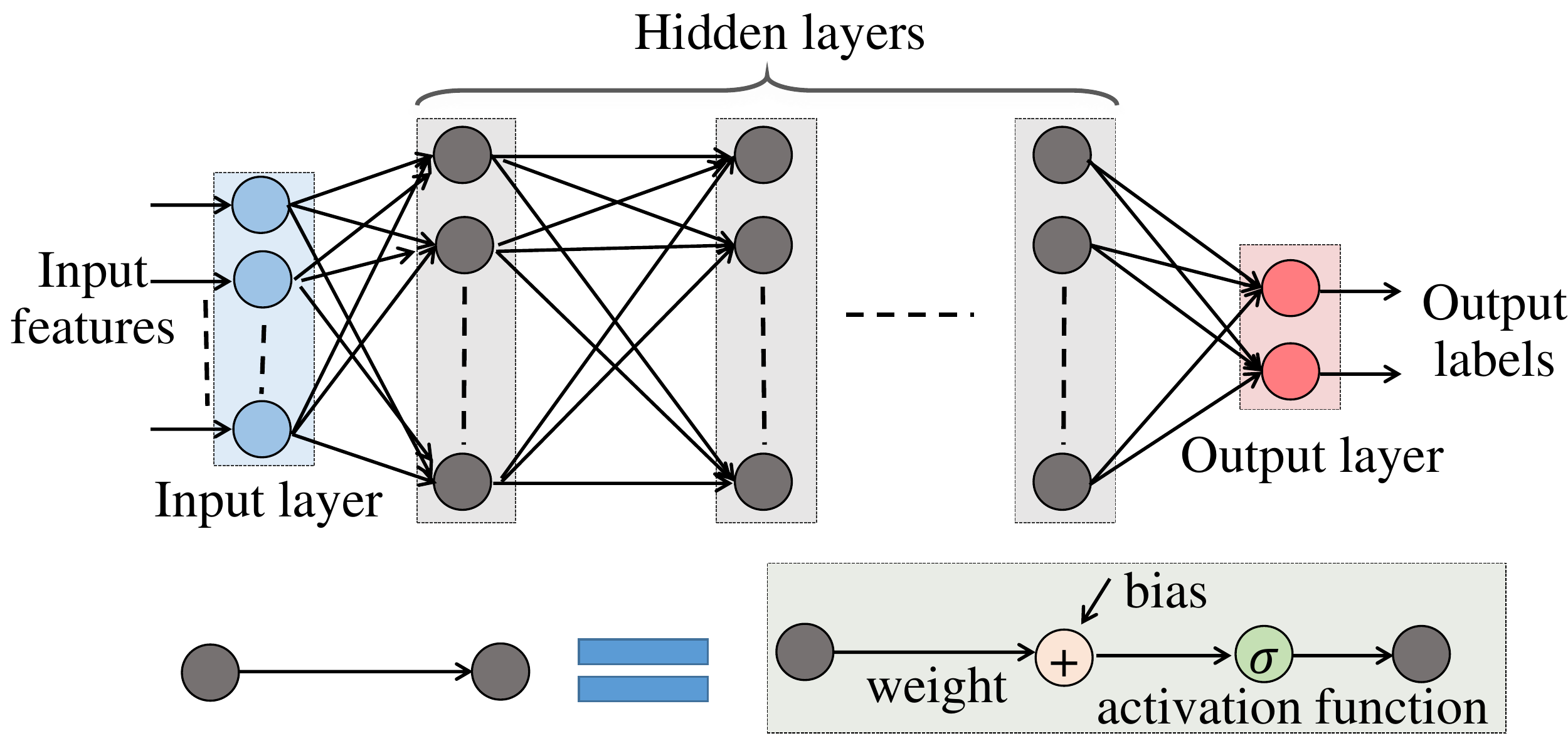}
	\caption{The structure of a feedforward neural network.}
	\label{fig:fnn}
\end{figure}

We use TensorFlow to build a deep learning classifier for $R$.
In particular, we use the following deep neural network:
\begin{itemize}
	\item A feedforward neural network is trained with backpropagation function by using cross-entropy as the loss function. The structure of a feedforward neural network is shown in Figure~\ref{fig:fnn}.
	\item Number of hidden layers is 3.
	\item Number of neurons per hidden layer is 50.
	\item Rectified linear unit (ReLU) is used as activation function at hidden layers.
	\item Softmax is used as the activation function at output layer.
	\item Batch size is 100.
	\item Number of training steps is 1000.
\end{itemize}
Note that $R$ can further optimize the hyperparameters (e.g., number of layers and number of neurons per layer) of its deep neural network.

In the simulation setting, the location of $T$ is $(0,0)$, the location of $R$ is $(10,0)$, the location of $A_T$ is $(0,10)$, the location of $A_R$ is $(10,0.1)$ (see Figs.~\ref{fig:topology1} and \ref{fig:topology2}), and the normalized transmit power at $B$ is $1000$.
$R$ collects $1000$ samples, each with $400$ spectrum sensing results, and a label (`$T$' or `not $T$') as training data and applies the classifier on another $1000$ samples to evaluate accuracy.
We used the above deep neural network and tuned its parameters.
We find that batch size $150$ can achieve better performance, while other parameters are unchanged.
There are $504$ signals from $T$ and $496$ signals from other transmitters in the test data.
Among them, $39$ signals from other transmitters are identified as from $T$ and $37$ signals from $T$ are identified as not from $T$.
Thus, we obtain $e_{FA} = 39/496 = 7.86\%$, $e_{MD} = 37/504 = 7.34\%$.
Both errors are small, showing that $R$ can reliably determine the signal labels.
We can also regard this case as a naive spoofing attack, where the adversary uses random signals to attack.
The success probability of this attack is only $7.86\%$.

\section{Spoofing Attacks}
\label{sec:adversary}

There is an adversary transmitter $A_T$ that aims to mimic the transmitter $T$'s behavior such that $R$ classifies signals from $A_T$ as from $T$.
$A_T$ does not have knowledge on $T$'s device related phase shift, or the channel between $T$ and $R$.
Instead, an adversary receiver $A_R$ is placed close to $R$ to learn received signal pattern from $T$ and from $A_T$. In the replay attack, $A_R$ is not used.

\subsection{Replay Attack based on Amplifying and Forwarding Signals}

We start with the replay attack based on simply amplifying and forwarding signals, i.e., $A_T$ records previously received signals from $T$, amplifies to power $P$ and forwards them to $R$.
Assume that $\theta_{A_T}$ is the phase shift for $A_T$, $\theta_{ij}$ is the phase shift and $g_{ij}$ is a random channel gain for the Rayleigh channel from node $i$ to node $j$. $T$'s parameters are unknown to $A_T$.
As an example, for two bits $0$ and $0$, QPSK determines a phase shift $\frac{\pi}{4}$.
the received phase shift at $R$ is $\frac{\pi}{4} + \theta_{TA_T} + \theta_{A_T} + \theta_{A_T R}$ and the received power is $g_{A_T R} P$.
 The $k$-th sampled data then becomes
\begin{eqnarray}
d_{A_T R}^k = g_{A_TR} P e^{j (\frac{\pi}{4} + \theta_T + \theta_{TA_T} + \theta_{A_T} + \theta_{A_T R})} \; \label{eq:dar}.
\end{eqnarray}
On the other hand, if the same signal is transmitted by $T$, the $k$-th sampled data is
\begin{eqnarray}
d_{TR}^k = g_{TR} P e^{j (\frac{\pi}{4} + \theta_T + \theta_{TR})} \; \label{eq:dtr}.
\end{eqnarray}
We can see that both received power and phase shift in (\ref{eq:dar}) and (\ref{eq:dtr}) are different.
However, a simple detector based on discriminating received power and/or phase shift cannot work.
The reason is that a detector needs to make a classification based on limited data samples, e.g., when an intruder is authenticated at the physical layer.
Due to random channel, the estimation of received power and phase shift cannot be accurate.
As a result, if we set a small region around the actual value of $T$, we will end up with a large misdetection probability.
On the other hand, if we set a large region, the false alarm probability will be large.

Simulation results show that the replay attack is better than transmitting random signals, i.e., the success probability of spoofing is increased to $36.2\%$, which is much larger than the success probability $7.86\%$ if $A_T$ transmits random signals.
This is because by amplifying and forwarding signals, only some signal pattern from $T$ is kept.
However, as we showed above, signals from $T$ and forwarded by $A_T$ are different even for the same data.
Thus, $R$ can still successfully classify most of spoofing signals.

\subsection{GAN-based Spoofing Attack}

We now consider the spoofing attack based on the GAN \cite{Goodfellow2014}.
As the first step, $A_R$ collects $500$ signal samples from $T$ and $500$ signal samples from $A_T$, where $A_T$ can flag its transmissions such that $A_R$ can have ground truth.
Each sample has coded data of $8$ bits under the QPSK modulation, i.e., $4$ signals.
The sampling rate for a signal is $100$, and thus the total data (features) for a sample is $400$.
$A_R$ then builds the first version of discriminator $D$ based on these data samples with the objective of minimizing classification error, i.e.,
\begin{eqnarray}
\min_{D} \mathbb{E}_{\bm{z} \sim p_{\bm{z}}} [\log(1 - D(G(\bm{z})))] - \mathbb{E}_{\bm{x} \sim p_{data}} [\log(D(\bm{x}))] \; ,
\end{eqnarray}
where $\bm{z}$ is a noise input to generator $G$ with a random distribution of $p_{\bm{z}}$ and $G(\bm{z})$ is the generator output and input data $\bm{x}$ has distribution $p_{data}$. 

On the other hand, $A_T$ collects the classification result from $A_R$, builds the first version of generator $G$ to generate synthetic data, and then transmits synthetic data to $A_R$.
The objective of $A_T$ is maximizing $A_R$'s classification error, i.e.,
\begin{eqnarray}
\max_{G} \mathbb{E}_{\bm{z} \sim p_{\bm{z}}} [\log(1 - D(G(\bm{z})))] - \mathbb{E}_{\bm{x} \sim p_{data}} [\log(D(\bm{x}))],
\label{eq:obj1}
\end{eqnarray}
where $D$ is the first version of discriminator.

This process continues with updated $G$ and $D$.
Both $G$ and $D$ have three hidden dense layers, each with $128$ neurons.
They improve in each round until convergence.
The entire process forms a GAN with a minimax game played between $A_T$ and $A_R$.
\begin{eqnarray}
\max_{G} \min_{D} \mathbb{E}_{\bm{z} \sim p_{\bm{z}}} [\log(1 - D(G(\bm{z})))] - \mathbb{E}_{\bm{x} \sim p_{data}} [\log(D(\bm{x}))] \; ,
\end{eqnarray}
although traditionally GAN is running at one entity.
Note that when $G$ is trained with the objective in (\ref{eq:obj1}), the gradients of $G$ rapidly vanish, which makes the training of GAN very difficult. To address the vanishing gradient problem, we use
\begin{eqnarray}
\max_{G} \mathbb{E}_{\bm{z} \sim p_{\bm{z}}} [\log(1 - D(G(\bm{z})))]
\end{eqnarray}
as the objective function at $G$ \cite{Goodfellow2014}.
Once converged, $A_T$ can apply its generator to generate and transmit synthetic signals.
After going through the wireless channel, the transmitted signals are received by $R$, which are statistically similar to the received signals from $T$.

We use the same simulation setting as that in Section~\ref{sec:classifier}.
The only difference is that signals from other transmitters is replaced by synthetic signals from $A_T$, where these synthetic signals are generated by $A_T$'s generator. For convergence, we check the maximum perturbation in $G$ and $D$ loss functions over the most recent 100 epochs of GAN training. When this perturbation drops below $5\%$ of current loss value, we terminate the GAN training. This way, the GAN is run only for $478$ epochs in this simulation setting. We find that the same classifier pre-trained at $R$, which works very well to discriminate signals from $T$ from random or replayed signals, cannot successfully identify synthetic signals generated by the GAN. The success probability of spoofing attack is increased to $76.2\%$, when the GAN is used by the adversary.

\begin{table}
\caption{Success probability of spoofing attack by different methods.}
\label{table:spoofing}
\small
\centering
\begin{tabular}{c|c}
\toprule
Method of spoofing attack & Success probability \\ \hline \hline
Random signal & $7.89\%$ \\ \hline
Replay (Amplify and forward) & $36.2\%$ \\ \hline
GAN-based spoofing & $76.2\%$ \\
\bottomrule 
\end{tabular}
\end{table}

Finally, we consider a more challenging case where $A_T$'s location is changed after the training process, e.g., $A_T$ moves from $(0,10)$.
$A_T$ may request the collaboration of $A_R$ to retrain a GAN and use its updated generator for spoofing attack.
If such update is not available, $A_T$ can still use its current generator to launch the attack.
Table~\ref{table:mobility} shows results under different $A_T$ locations. We can see that as $A_T$ moves away from $T$, the attack success probability decreases. This is an expected result since the distribution of the received channel characteristics varies as the receiver moves to a different location. However, the attack success probability is still significantly higher than the one achieved by the replay attack.

\begin{table}
\caption{The impact of $A_T$'s mobility on success probability of spoofing attack.}
\label{table:mobility}
\small
\centering
\begin{tabular}{c|c} \toprule
$A_T$'s location & Success probability \\ \hline \hline
$(0,10)$ & $76.2\%$ \\ \hline
$(0,11)$ & $65.2\%$ \\ \hline
$(0,15)$ & $61.0\%$ \\ \hline
$(0,20)$ & $56.2\%$ \\ \bottomrule
\end{tabular}
\end{table}

\section{Conclusion}
\label{sec:conclusion}

We designed a novel approach of spoofing wireless signals by generating synthetic signals by the GAN. We considered the case that an adversary transmits synthetic signals such that they are misclassified as the intended ones. We first showed that if there is no attack, a pre-trained deep learning-based classifier can distinguish signals reliably. We then considered a simple spoofing mechanism such as the replay (amplify-and-forward) attack that can only keep some pattern of intended signals. Therefore, the success probability of replay attack against a deep learning-based classifier remains limited. In this paper, we designed a GAN-based spoofing attack that generates synthetic data that is transmitted by an adversary transmitter and  distinguishes real and synthetic data at an adversary receiver. The minimax game between the adversary transmitter and receiver tunes both the generator and the discriminator. Then the signals generated by the GAN generator are transmitted for spoofing attack.
The GAN-based spoofing attack provides a major improvement in attack success probability over the random signal and replay attacks even when the node locations change from training to test time. As the GAN opens us new opportunities to effectively spoof wireless signals, new defense mechanisms are called for as future work.

\section*{Acknowledgments}

 This effort is supported by the U.S. Army Research Office under contract
W911NF-17-C-0090. The content of the information does not necessarily
reflect the position or the policy of the U.S. Government, and no official
endorsement should be inferred.

\end{document}